\documentclass[twocolumn,floatfix,preprintnumbers,superscriptaddress]{revtex4-1}

\usepackage[utf8]{inputenc}
\usepackage[colorlinks=true,citecolor=blue,linkcolor=blue]{hyperref}
\usepackage[normalem]{ulem}
\usepackage{url}
\usepackage{graphicx,wrapfig,float,slashed,cancel}
\usepackage{amsmath,amssymb,epsfig,graphicx,xcolor,stmaryrd}
\usepackage{bm}
\usepackage{enumitem}
\usepackage{subfigure}

\usepackage{amsfonts}
\usepackage{amssymb}
\usepackage{dsfont}
\usepackage{multirow}
\usepackage{appendix}
\usepackage{slashed}
\usepackage[active]{srcltx}
\usepackage{psfrag}
\usepackage{multirow}
\input epsf
\definecolor{darkblue}{RGB}{1, 90, 173}

\begin{document}   

\title{Analysis of the resonance X(4630) at non-zero temperature}
\date{\today}

\author{G. Bozk{\i}r }
\thanks{gbozkir@msu.edu.tr}
\affiliation{Department of Basic Sciences,  Army NCO Vocational HE School, National Defense University, Alt{\i}eyl\"{u}l, 
10185 Bal{\i}kesir,T\"{u}rkiye}

\begin{abstract}
We calculate the spectroscopic parameters of resonance $X(4630)$ observed in the process $B^{+}\rightarrow J/\Psi \phi K^{+}$ by at the LHCb experiments at CERN  by means of thermal QCD sum rule method at non-zero temperature. The exotic vector $X(4630)$ is assigned as the diquark-antidiquark state $[cs][\overline{cs}]$ with spin-parity $J^{PC}=1^{-+}$. Employing the two-point QCD sum rule approach up to the sixth order of the operator dimension by including non-perturbative contribution, we calculate the mass and decay constant of $X(4630)$ at $T\neq0$. The numerical analyses demonstrate that the values of the mass and decay constant of $X(4630)$ near the deconfinement temperature decrease up to $9.8\%$ and $60\%$ of their vacuum values. At $T\rightarrow0$, the obtained results for the mass $m_{X(4630)}=(4649\pm40)$ MeV and decay constant $\lambda_{X(4630)}=(10.07\pm0.8)\times10^{-3} $ MeV are in excellent agreement with the results reported by LHCb experiments and other theoretical predictions.

\end{abstract}

\maketitle
  

\section{INTRODUCTION}  \label{sec:intro}

The existence of quarks constructing hadrons and the quark model were first independently proposed by Gell-Mann \cite{Gell-Mann} and Zweig \cite{Zweig1, Zweig2} sixty years ago. In the quark model, a hadron is composed of one quark and one antiquark $q\overline{q}$ (called meson) or three quarks $qqq$ (called baryon). Besides, according to this model, there are hadrons apart from known mesons and baryons, called exotic hadrons, that are more quarks quark than mesons and baryons. The investigation of tetraquarks composed of $q\overline{q}q\overline{q}$ among exotic hadrons has been one of the fascinating subjects in particle physics since it has opened up a new perspective of hadron physics research, and this research on exotic hadrons has been started from the first observed of exotic charmonium $X(3872)$ by Belle Collaboration \cite{Choi1}. Besides, the observation of the X(3872) was reported later both in the CDF Collaboration \cite{Acosta} and D0 Collaboration by $\overline{p}p$ collisions \cite{Abazov} as well as in the CMS collaboration by $PbPb$ collisions \cite{Sirunyan}. After the discovery of $X(3872)$, many tetraquark states have been announced by both different theoretical approaches (such as the quark-bag model and QCD sum rules) \cite{Jaffe1, Jaffe2, Lloyd, Navarra, Matheus, Lee1, Chen1, Lu, Becchi1, Becchi2, Albuquerque, Zhang, Yang1, Wang1, Faustov, Kuang, Dong, Niu, Agaev1, Agaev2, Agaev3, Agaev4, Agaev5, Agaev6, Agaev7} and by heavy ion collision experiments \cite{Chen2, Aaij1, Aaij3, Aaij4, Aaij5}.

In 2021, the charmonium-like resonances such as $Z_{cs}(4000)^{+}$, $Z_{cs}(4220)^{+}$, $X(4630)$ and $X(4685)$ were observed in the process $B^{+}\rightarrow J/\Psi \phi K^{+}$ with proton-proton collision by the LHCb Collaboration at CERN \cite{Aaij2}. In this experiment, $X(4630)$ and $X(4685)$ states were observed in the $J/\Psi \phi$ decay channel as $Z_{cs}(4000)^{+}$ and $Z_{cs}(4220)^{+}$ states were reported in $J/\Psi K^{+}$ channel. The measured mass and width of $X(4630)$ resonance with $[cs][\overline{cs}]$ diquark-antidiquark state are 

\begin{eqnarray}\label{massandwidthexp}
m_{X(4630)}=(4626\pm 16^{+18}_{-110})MeV,\notag \\ 
\Gamma_{X(4630)}=(174\pm 27^{+134}_{-73})MeV.
\end{eqnarray}

On the other hand, we know that theoretically predicting the behavior of the X(4630) charmonium-like resonance plays a crucial role in interpreting the experimental results. So far, a few studies in the literature are dedicated to this aim \cite{Liu, Yang2,Wang2,Agaev8,GangYang}. In \cite{Liu}, $[cs][\overline{cs}]$ diquark-antidiquark states with $J^{PC}=0^{++},1^{++},1^{+-},2^{++}$ have been investigated using the quark delocalization color screening model and energy ranges for any resonances states have been obtained. Yang et al have proposed that the detected $X(4630)$ resonance is a candidate of $D_{s}^{*}\overline{D}_{s1}(2536)$ charmonium-like molecule with $J^{PC}=1^{-+}$ in the framework of the one-boson-exchange mechanism in \cite{Yang2}. In $X(4630) \rightarrow J/\Psi \phi$ decay channel, they have noticed that the spin-parity of $X(4630)$ resonance must be $J^{PC}=1^{-+}$ due to the conservation of the C-parity and these state can not be interpreted as a conventional meson. In \cite{Wang2}, Wang has assigned the $X(4630)$ resonance as the $D_{s}^{*}\overline{D}_{s1}(2536)$ charmonium-like molecule in the mass spectrum of $J/\Psi \phi$ decay channel via the QCD sum rules as agree with the \cite{Yang2}. In \cite{Agaev8}, the mass and width of the $X(4630)$ resonance in vacuum have been calculated within the QCD sum rules.  In \cite{GangYang}, all possible charmonium-like tetraquark states with the $I(J^{P})=0(0^{+}),0(1^{+}),0(2^{+}),1(0^{+}),1(1^{+}),1(2^{+})$ have been investigated using the meson–meson, diquark–antidiquark and K type configurations in a chiral quark model.
All of these studies have been performed in vacuum. However, in recent years, the investigation of exotic charmonium states under high temperature and/or dense medium has aroused the interest of theoretical physicists to deeply understand the nature of the hadronization phase of the quark-gluon plasma (QGP). It is known that a phase transition around critical temperature $T_c=155$ MeV may occur from hadron gas (confined matter) to QGP which is accepted as a new state of matter. QGP is formed in the initial stage of the heavy ion collision (HIC) experiment and it is expected that hadronization occurs by the temperature decreasing from $300-400$ MeV to $T_c=155$ MeV. The abundance of all exotic states detected in the hadron phase after the QGP has hadronized in HIC can change due to charm and bottom quarks created at the QGP phase interacting with the light hadrons such as pions, kaons, and rhos in the medium through both elastic and inelastic processes.  These hadronic effects on the abundance of exotic particles have been discussed using an effective Lagrangian approach in detail \cite{Lee2, Torres, Abreu1, Britto, Abreu2}. Also,  these interactions are described by condensates which contains masses of the exotic states, and all condensates  vanish in QGP and replace with their thermal condensates. Therefore, investigating of the temperature effects on the spectroscopic parameters such as mass and decay constant of exotic particles is important to understand the chiral phase transitions, the restoration of chiral symmetry, and the properties of hot-dense matter in QCD.

In the last decades, the modifications of the masses and decay constants of $X(3872)$ \cite{Veliev}, $Y_{b}(10890)$ \cite{Sungu1}, $X(5568)$ \cite{Sungu2}, $Z_{cs}(3985)$ \cite{Sungu3} exotic states at finite temperature have been investigated using the thermal QCD sum rules approach. In \cite{Veliev, Sungu1, Sungu2, Sungu3}, the masses and decay constants of these exotic states almost remain unchanged up to $T\cong100-120$ MeV but, they begin decrease exponentially by increasing the temperature after this point. In \cite{Montana}, properties of $X(3872), X(4014)$ exotic particles at finite temperature have been investigated using an effective hadron theory and obtained results show that the masses of them would drop $\Delta m \cong - 60$ MeV, and acquire a decay width of $\Gamma\cong120 MeV$. In \cite{Montesinos}, the modifications of the masses and the widths of the tetraquark-like $T_{cc}(3875)^{+}$ and $T_{\overline{c} \overline{c}}(3875)^{-}$ states in dense nuclear matter have been investigated using an effective hadron approach. It has been shown that the  $T_{cc}(3875)^{+}$ in matter becomes broader than the $T_{\overline{c} \overline{c}}(3875)^{-}$, whereas the mass of the former is moved to larger values than the nominal mass and the mass of the latter is displaced to smaller ones and their widths grow with increasing density.

In the present study, our aim  to investigate the thermal behavior of the exotic $X(4630)$ resonance at non-zero temperature. Our calculations on the mass and decay constant of $X(4630)$ are performed in the thermal QCD sum rule method framework, and their temperature-dependent values are discussed. The thermal QCD sum rules method is the thermal version of QCD sum rules first introduced in the 1970s by Shifman, Vainshtein, and Zakharov \cite{Shifman}, and accepted as one of the most potent non-perturbative methods. In the thermal QCD sum rules, the calculations are performed taking into account energy-momentum tensor including fermionic and gluonic parts as well as the temperature-dependent expressions of the quark, gluon, and mixed condensates \cite{Bochkarev}.

This article is organized as follows: In Section~\ref{II}, the thermal QCD sum rules derived to compute for the mass and decay constant of the $X(4630)$ resonance at non-zero temperature are given. Section~\ref{III} is devoted to presenting the numerical analyses of the obtained thermal QCD sum rules in the previous section. In Section~\ref{IV}, the summary and conclusions are presented.

\section{THE THERMAL QCD SUM RULES FOR THE MASS AND DECAY CONSTANT}\label{II}

In this section, our aim is to obtain the thermal QCD sum rules for the mass and decay constant of the $X(4630)$ resonance at non-zero temperature. As a starting point, we need to construct the following two-point thermal correlation function

\begin{equation}\label{CorrFuncAxial}
\Pi_{\mu \nu }(q,T)=i\int d^{4}x~e^{iq\cdot x}\langle
\Psi|\mathcal{T}\Big\{\eta_{\mu}(x)\eta_{\nu}^{\dagger}(0)\Big\}|\Psi\rangle,
\end{equation}
where $q$ indicates the four-momentum of the $X(4630)$ state, $\mathcal{T}$ is the time-ordering operator for currents and $\Psi$ represents the ground state operator of hot medium.  For the $X(4630)$ resonance with spin-parity $J^{PC}=1^{-+}$, the interpolating current $\eta_{\mu}(x)$ is given as 

\begin{eqnarray}\label{TetraCurrent}
\eta_{\mu }(x)&=&\frac{i\epsilon_{abc} \epsilon_{dec}}{\sqrt{2}}\Big\{
\Big[~s_{a}^{T}(x)C\gamma_{5} c_{b}(x)\Big] \Big[~
\overline{s}_{d}(x)\gamma_{\mu}\gamma_{5}C\overline{c}_{e}^{T}(x)\Big]  \nonumber \\
&+&\Big[~s_{a}^{T}(x)C\gamma_{5}\gamma_{\mu}c_{b}(x)\Big] \Big[~
\overline{s}_{d}(x)\gamma_{5}C\overline{c}_{e}^{T}(x)\Big]\Big\}.
\end{eqnarray}
In Eq.~(\ref{TetraCurrent}), $C$ represents charge conjugation matrix, $\epsilon_{abc}$ and $\epsilon_{dec}$ are anti-symmetric Levi-Civita tensors that $a,b,c,d,e$ denotes color indices. 

The second step of the derivation of the thermal QCD sum rules is that the two-point thermal correlation function given in Eq.~(\ref{CorrFuncAxial}) is calculated from two different paths: i) the physical side evaluated in terms of the temperature-dependent values of the physical parameters such as mass and decay constant of considered hadron in the time-like region, ii) the OPE side evaluated by Operator Product Expansion (OPE) in terms of the quark and gluon fields in hot medium.

To obtain the physical side of the correlation function, we insert the interpolating current with the same quantum numbers given in Eq.~(\ref{TetraCurrent}) into the above correlation function and perform the four-integral over $x$ in Eq.~(\ref{CorrFuncAxial}). As a result, we get the physical side of the correlation function as 

\begin{equation}\label{Physicalcorrelator}
\Pi _{\mu \nu }^{\mathrm{Phys}}(q,T)=\frac{\langle \Psi|\eta_{\mu
}|X(q)  \rangle_T \langle X(q)  |\eta_{\nu }^{\dagger
}|\Psi\rangle_T }{(m_{X(4630)}^{2}(T)-q^{2})}
+ST.
\end{equation}

Here, $m_{X(4630)}(T)$ is the temperature-dependent mass of $X(4630)$
resonance and ST represent the subtracted terms involving the contributions of the higher resonances and continuum states. The matrix element for $X(4630)$ vector state in terms of the temperature-dependent decay constant $\lambda_{X(4630)}(T)$ is introduced as

\begin{equation}\label{Polarizationvector}
\langle \Psi|\eta_{\mu }|X(q)  \rangle_T
=\lambda_{X(4630)}(T) m_{X(4630)}(T)~\varepsilon _{\mu},
\end{equation}
where $\varepsilon _{\mu}$ is the polarization vector of the $X(4630)$ having a relationship given as $\varepsilon_{\mu}\varepsilon_{\nu}^{*}=-g_{\mu\nu}+q_{\mu}q_{\nu }/m_{X(4630)}^{2}(T)$. Taking into account the above-introduced matrix element in Eq.~(\ref{Physicalcorrelator}), we reform the physical side of the correlation function in terms of different Lorentz structures as 

\begin{eqnarray}\label{CorM}
\Pi _{\mu \nu }^{\mathrm{Phys}}(q,T)&=&\frac{m_{X(4630)}^{2}(T)\lambda_{X(4630)}^{2}(T)} {%
(m_{X(4630)}^{2}(T)-q^{2})}\nonumber \\
&\times &\left( -g_{\mu \nu }+\frac{q_{\mu }q_{\nu }}{m_{X(4630)}^{2}(T)%
}\right)+ST.
\end{eqnarray}
To calculate the physical and OPE sides of the correlation function in this study, we choose coefficients of the structure $g_{\mu \nu}$. It is known that choosing the structure $g_{\mu \nu}$ for spin-1 states is advantaged due to this structure receives a contribution from only vector particles, on the other hand, the structure  $q_{\mu}q_{\nu }$ contains both spin-0 and spin-1 states. To remove ST terms arising in correlation functions in thermal QCD sum rule calculations, we apply the standard Borel transformation with respect to $q^{2}$. After applying the Borel transformation to correlation functions and equating coefficients of $g_{\mu \nu}$ terms for two sides of the correlation function, thermal QCD sum rules for the mass and decay constant of the $X(4630)$ resonance at non-zero temperature are obtained. Also, we will analyze the Borel parameter $M^2$ and continuum subtraction parameter $s_{0}$ arising in the Borel transformation in numerical computations. By using these procedures, the final expression of the physical side is written as

\begin{eqnarray}\label{CorBor}
\mathcal{B}(q^2)\Pi^{\mathrm{Phys}}(q,T)&=& m_{X(4630)}^{2}(T)\lambda_{X(4630)}^{2}(T)\nonumber \\
&\times & e^{-m_{X(4630)}^{2}(T)/M^{2}}.
\end{eqnarray}
On the OPE side of the correlation function for $X(4630)=[cs][\overline{cs}]$, we insert the interpolating current $\eta_{\mu}(x)$ in Eq.~(\ref{CorrFuncAxial}) and obtain the OPE side of the correlation function in terms of the heavy $c$ quark propagator $S_{c}^{ij}(x)$ and light $s$ quark propagator $S_{s}^{ij}(x)$ by contracting between quark fields: 

\begin{eqnarray}
\Pi _{\mu \nu }^{\mathrm{OPE}}(q,T)&=&-\frac{i}{2}\int
d^{4}xe^{iq\cdot x}\epsilon^{abc} \epsilon^{dec}\epsilon^{a^{'}b^{'}c^{'}} \epsilon^{d^{'}e^{'}c^{'}}\nonumber \\
&\times& \langle\{\mathrm{Tr}[\gamma_{\mu}\gamma_{5}\widetilde{S}_{c}^{aa^{\prime}}(-x)\gamma_{5}\gamma_{\nu}S_{s}^{bb^{\prime}}(-x)]\nonumber \\
&\times&\mathrm{Tr}[\gamma_{5}\widetilde{S}_{s}^{dd^{\prime}}(x)\gamma_{5}S_{c}^{ee^{\prime}}(x)]\nonumber \\
&+&\mathrm{Tr}[\gamma_{5}\widetilde{S}_{c}^{aa^{\prime}}(-x)\gamma_{5}S_{s}^{bb^{\prime}}(-x)\gamma_{\mu}]\nonumber \\
&\times&\mathrm{Tr}[\gamma_{5}\widetilde{S}_{s}^{dd^{\prime}}(x)\gamma_{5}\widetilde{S}_{c}^{ee^{\prime}}(x)
\gamma_{\nu}\gamma_{5}S_{c}^{bb^{\prime}}(x)]\nonumber \\
&+&\mathrm{Tr}[\gamma_{5}\widetilde{S}_{c}^{aa^{\prime
}}(-x)\gamma_{5}\gamma_{\nu}S_{s}^{bb^{\prime}}(-x)]
\nonumber \\
&\times&\mathrm{Tr}[\gamma_{5}\widetilde{S}_{s}^{dd^{\prime}}(x)\gamma_{5}\gamma
_{\mu}S_{c}^{ee^{\prime}}(x)]\nonumber \\
&+&\mathrm{Tr}[\gamma_{5}\widetilde{S}_{c}^{aa^{\prime
}}(-x)\gamma _{5}S_{s}^{bb^{\prime }}(-x)]\nonumber \\
&\times&\mathrm{Tr}[ \gamma
_{5}\widetilde{S}_{s}^{dd^{\prime }}(-x)\gamma
_{5}S_{c}^{ee^{\prime }}(x)\gamma_{\nu}]\} \rangle_{T}.
\end{eqnarray}
Here, $\widetilde{S}_{s(c)}^{ij}(x)$, $S_{s}^{ij}(x)$ and $S_{c}^{ij}(x)$ in coordinate space are given by the following expressions:
\begin{equation}\label{tildapropagator}
\widetilde{S}_{s(c)}^{ij}(x)=CS_{s(c)}^{ijT}(x)~C,
\end{equation}
\begin{eqnarray} \label{LQprop}
S_{s}^{ij}(x)
&=&i\frac{{\!\not\!{x}}}{2\pi^{2}x^{4}}\delta_{ij}-\frac{m_s}{4\pi^{2}x^2}\delta_{ij}
-\frac{\langle\bar{s}s\rangle}{12}\delta_{ij}\nonumber\\
&-&\frac{x^2}{192}m_{0}^{2}\langle
\bar{s}s\rangle \Big[1-i\frac{m_{s}}{6}{\!\not\!{x}}
\Big]\delta_{ij}\nonumber\\
&+&\frac{i}{3}\Big[{\!\not\!{x}}\Big(\frac{m_{s}}{16}\langle
\bar{s}s\rangle -\frac{1}{12}\langle u\Theta ^{f}u\rangle \Big)\nonumber\\
&+&\frac{1}{3}\Big(u\cdot x\Big){\!\not\!{u}} \langle u\Theta^{f}u\rangle %
\Big]\delta _{ij} \nonumber\\
&-&\frac{ig_{s}\lambda_{ij}^{A}}{32\pi ^{2}x^{2}}G_{A}^{\mu \nu }
\Big({\!\not\!{x}} \sigma _{\mu \nu }+\sigma _{\mu \nu
}{\!\not\!{x}}\Big),
\end{eqnarray}
\begin{eqnarray}\label{HeavyProp}
S_{c}^{ij}(x)&=&i\int \frac{d^{4}k}{(2\pi )^{4}}e^{-ik\cdot x}\Bigg[ \frac{%
\delta _{ij}\Big( {\!\not\!{k}}+m_{c}\Big)
}{k^{2}-m_{c}^{2}}\nonumber\\
&-&\frac{gG_{ij}^{\alpha \beta }}{4}\frac{\sigma _{\alpha \beta }\Big( {%
\!\not\!{k}}+m_{c}\Big) +\Big(
{\!\not\!{k}}+m_{c}\Big)\sigma_{\alpha
\beta }}{(k^{2}-m_{c}^{2})^{2}}\nonumber \\
&+&\frac{g^{2}}{12}G_{\alpha \beta }^{A}G_{A}^{\alpha \beta
}\delta_{ij}m_{b}\frac{k^{2}+m_{c}{\!\not\!{k}}}{(k^{2}-m_{c}^{2})^{4}}+\ldots\Bigg].
\end{eqnarray}
In the above expressions, $m_{s(c)}$ is the mass of $s(c)$ quark, $\langle \bar{s}s\rangle $ implies the temperature-dependent of the $s$ quark condensate, $G_{ij}^{\alpha \beta}=G_{A}^{\alpha \beta}\lambda_{ij}^{A}/2$ is the external gluon field with the Gell-Mann matrices $\lambda_{ij}^{A}$ ($A=[1-8]$ is color index taking a value in the range $[1-8]$), $\Theta _{\mu \nu}^{f}$ is the fermionic part of the energy-momentum tensor and $u_{\mu }$ represents the four-velocity for the hot medium in the rest frame $(u_{\mu }=(1,0,0,0))$. For the trace of the two-gluon field strength tensor, we use the 
following expression presented in \cite{Mallik}   

\begin{eqnarray}\label{TrGG} 
\langle Tr^c G_{\alpha \beta} G_{\mu \nu}\rangle &=& \frac{1}{24} (g_{\alpha \mu} g_{\beta \nu} -g_{\alpha
	\nu} g_{\beta \mu})\langle G^2\rangle_{T} \nonumber \\
&+&\frac{1}{6}\Big[g_{\alpha \mu}g_{\beta \nu} -g_{\alpha \nu} g_{\beta \mu} -2(u_{\alpha} u_{\mu}g_{\beta \nu} \nonumber \\
&-&u_{\alpha} u_{\nu} g_{\beta \mu} -u_{\beta} u_{\mu}
g_{\alpha \nu} +u_{\beta} u_{\nu} g_{\alpha \mu})\Big]\nonumber \\
&\times&\langle u^{\lambda} {\Theta}^g _{\lambda \sigma} u^{\sigma}\rangle,
\end{eqnarray}
where $\Theta_{\lambda \sigma }^{g}$ is the gluonic part of
the energy-momentum tensor. $\Pi _{\mu \nu }^{\mathrm{OPE}}$ can be written in the following different Lorentz structures form:
\begin{eqnarray}
\Pi_{\mu \nu }^{\mathrm{OPE}}(q,T)&=&\Pi _{0}^{\mathrm{OPE}}(q^{2},T)\frac{%
q_{\mu }q_{\nu }}{q^{2}}\nonumber \\
&+&\Pi_{1}^{\mathrm{OPE}}(q^{2},T)(-g_{\mu \nu }+\frac{q_{\mu }q_{\nu }}{%
q^{2}}).
\end{eqnarray}
According to the thermal QCD sum rule procedure,  we should choose the $g_{\mu \nu}$ structure like in the physical side and continue the calculating using $\Pi _{1}^{\mathrm{OPE}}(q^{2},T)$. For this aim, we define the coefficient of $g_{\mu \nu}$ structure as 

\begin{equation}
\Gamma_{1}^{\mathrm{OPE}}(q^{2},T)=\Pi _{1}^{\mathrm{OPE}}(q^{2},T)g_{\mu\nu}.
\end{equation}
Thus, $\Gamma_{1}^{\mathrm{OPE}}(q^{2},T)$ is given with a dispersion relation as
\begin{equation}
\label{disp_0}
\Gamma^{OPE}_{1}(q^{2},T)=\int_{4(m_c+m_s)^2}^\infty ds \dfrac{\rho^{\mathrm{OPE}}(s,T)}{s-q^2},
\end{equation}
where the temperature-dependent spectral density $\rho^{\mathrm{OPE}}(s,T)$  is given as 
\begin{equation}
\rho^{\mathrm{OPE}}(s,T)=\frac{1}{\pi}\mathrm{Im}[\Gamma^{OPE}_{1}(s,T)].
\end{equation}
$\rho^{\mathrm{OPE}}(s,T)$ includes both perturbative and non-perturbative components (up to dimension six) and these components will be presented in Appendix A. After performing the Borel transformation and continuum subtraction on the obtained $\Gamma^{OPE}_{1}(s,T)$, we find 

\begin{eqnarray}\label{Gammafunc}
\hat{B}\Gamma_{1}^{OPE} (T)=\int_{4(m_c+m_s)^2}^{s_{0}(T)} ds \rho^{\mathrm{OPE}}(s,T)e^{-s/M^2},\nonumber\\
\end{eqnarray}
where $s_{0}(T) $ represents the temperature-dependent expression of continuum threshold.

Finally, by matching the coefficients of the OPE side and of the physical side of the correlation function by the quark hadron duality, we obtain thermal sum rules for the mass and decay constant of the $X(4630)$ resonance at non-zero temperature.  

\begin{equation}\label{ResidueSR}
\hat{B}\Gamma_{1}^{OPE} (T)=m_{X(4630)}^{2}(T)\lambda_{X(4630)}^{2}(T)~e^{-m_{X(4630)}^{2}(T)/M^{2}}.
\end{equation}
To extract the mass of the $X(4630)$ resonance, we take the derivative of the above-given sum rule with respect to $(-\frac{1}{M^{2}})$ and divide this derivative by itself:

\begin{equation}\label{massSR}
m_{X(4630)}^{2}(T)=\frac{\int_{4(m_{c}+m_{s})^{2}}^{s_{0}(T)}ds~s~\rho(s,T)~e^{-s/M^{2}}}{\int_{4(m_{c}+m_{s})^{2}}^{s_{0}(T)}ds~\rho(s,T)~e^{-s/M^{2}}}.
\end{equation}
%

\section{Numerical results}\label{III}

This section aims to present the numerical results obtained for the mass and decay constant of the $X(4630)$ resonance at zero and non-zero temperatures. For this purpose, we first collect vacuum values of the various input parameters used in the numerical calculations in Table \ref{tab:inputPar}.

\begin{table}[ht!] 
	\centering
	\begin{tabular}{ |c|c|}
		\hline \hline
		Parameters  &  Values   \\ \hline
		$m_{c}$   &  $ (1.27\pm0.02)~\mathrm{GeV} $    \\    
		$ m_{s}$  &  $ (93.4_{-3.4}^{+8.6})~\mathrm{MeV}$      \\  
		$m_{0}^{2}$   &  $(0.8\pm0.2)\mathrm{GeV}^2$  \\ 
		$ \langle0| s\bar{s}|0\rangle $  &  $-0.8\times(0.24\pm0.01)^3~\mathrm{GeV}^3 $  \\ 
		$ \langle0|\frac{1}{\pi}\alpha_sG^2|0\rangle$   &  $ (0.012\pm0.004)~\mathrm{GeV}^4$  \\ 
		\hline \hline
	\end{tabular}
	\caption{Vacuum values of the input parameters. \cite{Zyla,Belyaev1,Belyaev2,Ioffe}.}
	\label{tab:inputPar}
\end{table}

On the other hand, we need to know some thermal parameters such as the thermal quark and gluon condensates as well as the gluonic and fermionic parts of the energy density arising at $T\neq0$. For these parameters, we use the following expressions described in \cite{Azizi1,Ayala1,Ayala2} :

\begin{equation}\label{eq::qqT}
\langle \bar{q}q\rangle =\frac{\langle 0|\bar{q}q|0\rangle }{%
1+exp~\Bigg({\big(A\big[\frac{1}{\mathrm{\mathrm{GeV}}^{2}}\big]T^{2}+B\big[\frac{1}{%
\mathrm{\mathrm{GeV}}}\big]T-1\big)}\Bigg)},
\end{equation}
\begin{eqnarray}\label{G2TLattice}
\langle G^{2}\rangle  =\langle 0|G^{2}|0\rangle \Bigg[1-C\Bigg(\frac{T}{%
T_{c}}\Bigg)^{E}+D\Bigg(\frac{T}{T_{c}}\Bigg)^{F}\Bigg],
\end{eqnarray}
and
\begin{eqnarray}\label{tetamumu}
\langle \Theta _{00}^{g}\rangle &=&\langle \Theta^f
_{00}\rangle=T^{4}exp~{\Big(G\Big[\frac{1}{\mathrm{GeV}^{2}}\Big]T^{2}-H\Big[\frac{1}{\mathrm{GeV}}\Big]T
\Big)} \nonumber \\
&-& K\Big[\frac{1}{\mathrm{GeV}}\Big]T^{5}.
\end{eqnarray}
Here, $A=18.10042$, $B=4.99216$, $C=1.65$, $D=0.04967$, $E=8.735$, $F=0.7211$, $G=113.867$, $H=12.190$ and $J=10.141$. Also, $T_c$ is a critical temperature commented as the deconfinement temperature, and in our calculations we take it as $T_c=155$ MeV as assumed in \cite{Ayala2}.

\begin{figure}[ht]
	\begin{center}
		\includegraphics[width=7.5cm]{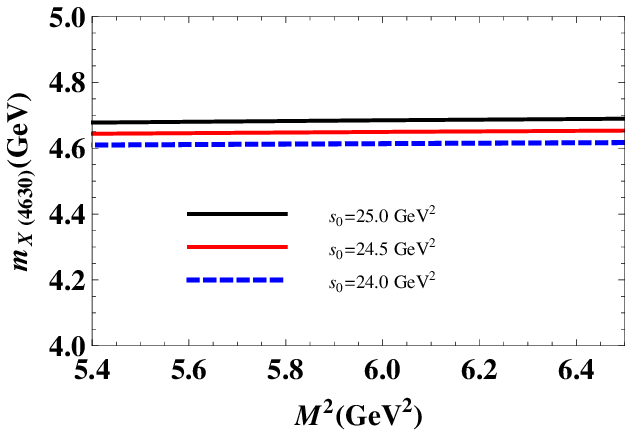}
	\end{center}
	\caption{For the values of the fixed continuum thresholds, the mass of $X(4630)$ resonance at $T=0$ according to the Borel parameter.} \label{fig1}
\end{figure}

\begin{figure}[ht] 
	\begin{center}
		\includegraphics[width=7.5cm]{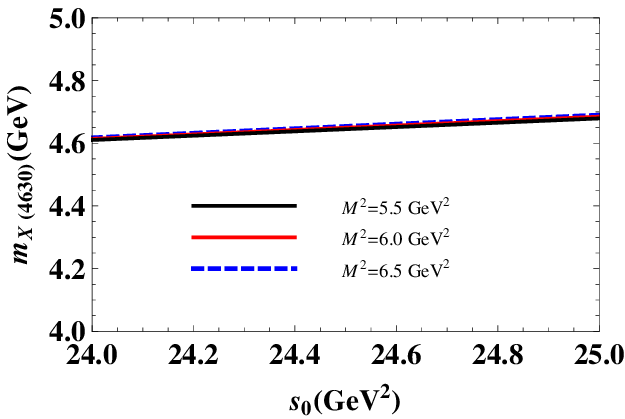}
	\end{center}
	\caption{For the values of the fixed Borel parameters, the mass of $X(4630)$ resonance at $T=0$ according to the continuum threshold.} \label{fig2}
\end{figure}

To proceed further in the numerical calculation, we need to determine the working regions of two auxiliary parameters entering the thermal QCD sum rule. These auxiliary parameters are the Borel parameter $M^{2}$ and the continuum threshold $s_0$, and they should be independent of physical quantities according to the thermal QCD sum rules procedure. For this aim, we plot the mass of the $X(4630)$ resonance in vacuum versus $M^{2}$ and $s_0$ in Figures \ref{fig1} and \ref{fig2}, respectively. We see that the mass is nearly stable in the following given intervals of $M^{2}$ and $s_0$:

\begin{eqnarray*}
5.5~\mathrm{GeV}^2\leq M^2 \leq 6.5~\mathrm{GeV}^2,
\end{eqnarray*}
and 
\begin{eqnarray*}
24~\mathrm{GeV}^2\leq s_0 \leq 25~\mathrm{GeV}^2.
\end{eqnarray*}

\begin{widetext}

\begin{figure}[h!]
\begin{center}
\includegraphics[totalheight=6cm,width=8cm]{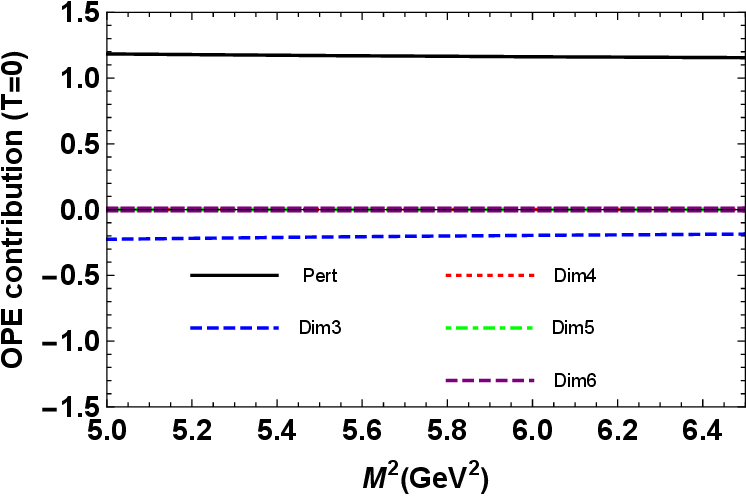}\,\, %
\includegraphics[totalheight=6cm,width=8cm]{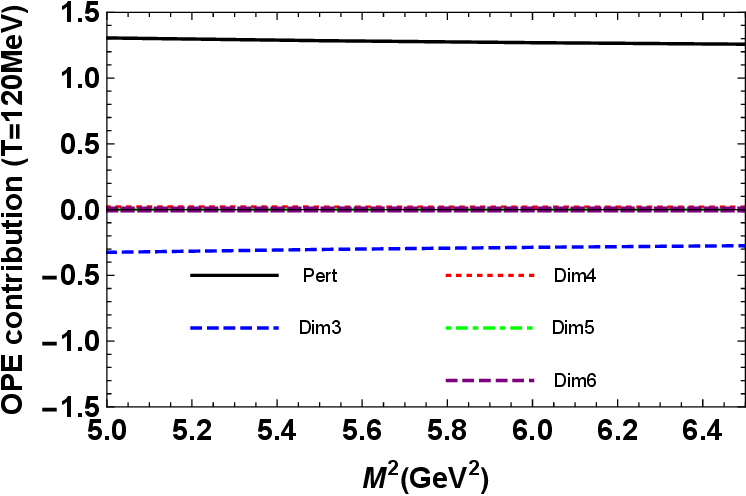}
\end{center}
\caption{For $s_{0}=24.5$, different dimensional OPE contributions at $T=0$ and  $T=120$MeV according to the Borel parameter. }
\label{fig3}
\end{figure}

\end{widetext}

Also, we need to check the convergence of OPE in the above-determined working regions for $M^{2}$ and $s_0$. The contributions of the highest dimensional operators in the OPE should be less than the contributions of all OPE terms entering the calculations. To test this, we can be written $R(M_{\mathrm{min}}^{2})$ ratio given bellow

\begin{equation}
R(M_{\mathrm{min}}^{2})=\frac{\Pi^{\mathrm{dimN}}(M_{\mathrm{%
min}}^{2},\ s_{0})}{\Pi(M_{\mathrm{min}}^{2},\ s_{0})}
\label{eq:Cond2}
\end{equation}
where $\Pi ^{\mathrm{dimN}}(M_{\mathrm{%
min}}^{2},s_{0})$ denotes the contribution of the operator with $N$ dimension in OPE. We calculated this ratio for both $T=0$ and $T=120MeV$ and we saw that the contribution of the highest-dimension operator does not exceed 0.001. For a more detailed analysis, we plot $R(M_{\mathrm{min}}^2)$ according to the Borel parameter at both $T=0$ and $T=120MeV$ in Figure \ref{fig3}. As a result, the OPE convergence is valid in our sum rules.

We employ the values that $s_0=24,24.5,25$ and $M^{2}=6$ for our computations considering the above-mentioned working intervals. Using these values for $M^{2}$ and $s_0$, we obtain the values of mass and decay constant of the $X(4630)$ resonance at $T=0$ and present the average of the obtained results with the results of the other works in Table \ref{tab:results}. The presented errors for our results in this table come from varying values in working  regions of the auxiliary parameters, $s_0$ and $M^{2}$.

\begin{table}[ht!] 
	\centering
	\begin{tabular}{|c|c|c|}
		\hline \hline 
	         & $m_{X(4630)}(MeV)$ &  $\lambda_{X(4630)}(GeV^4)$  \\ \hline
	Our results  &	$ (4649\pm40) $   &  $ (10.07\pm0.8)\times10^{-3} $   \\    
	Experiment \cite{Aaij2}   &	$(4626\pm 16^{+18}_{-110})$  &  -      \\  
	\cite{Yang2,Wang2}   &	 $(4630^{+110}_{-80}) $   &  -  \\ 
	 \cite{Agaev8}   &	$ (4632\pm60) $  &  $ (9.2\pm0.8)\times10^{-3} $  \\ 
		\hline \hline
	\end{tabular}
	\caption{The values of mass and decay constant of the $X(4630)$ resonance at $T=0$.}
	\label{tab:results}
\end{table}

\begin{figure}[ht] 
	\begin{center}
		\includegraphics[width=7.5cm]{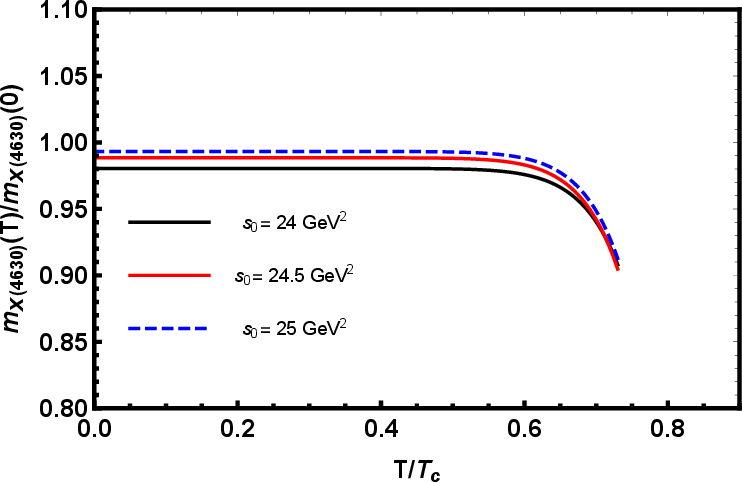}
	\end{center}
	\caption{The ratio $ m_{X(4630)}(T)/m_{X(4630)}(0)$ as a
function of $T/T_{c}$ at different values of the continuum threshold.} \label{fig4}
\end{figure}

\begin{figure}[ht] 
	\begin{center}
		\includegraphics[width=7.5cm]{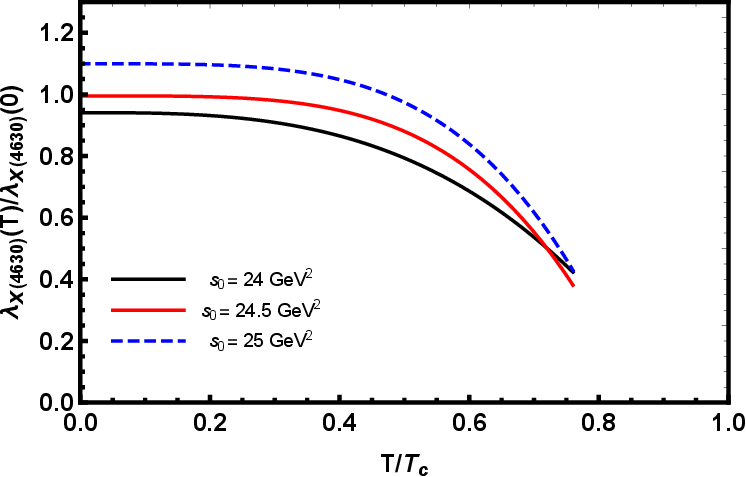}
	\end{center}
	\caption{The ratio $ \lambda_{X(4630)}(T)/\lambda_{X(4630)}(0)$ as a
function of $T/T_{c}$ at different values of the continuum threshold.}\label{fig5}
\end{figure}

Considering the suitable working intervals given above for $M^{2}$ and $s_0$, we obtain the numerical values for the mass and decay constant of the $X(4630)$ resonance in the hot medium. To search the dependence of the mass and decay constant of $X(4630)$ on the temperature, we plot the ratio of thermal values of these physical quantities to their vacuum values in Figures \ref{fig4} and \ref{fig5}. As is seen from these figures, the mass and decay constant stay approximately non-changed up to $T/T_{c}\simeq0.5$ and $T/T_{c}\simeq0.3$, respectively. However, after these points, the mass decreases by \textit{$9.8\%$}  of its vacuum value near $T/T_{c} \simeq 0.7$ as the decay constant decreases by \textit{$60\%$} of its vacuum value near $T/T_{c} \simeq 0.8$. The decreasing of the mass and decay constant by increasing temperature can be commented on as the QGP phase transition. Also, these results will be checked using future heavy ion collision experiment data.
%

\section{Summary and Conclusions}\label{IV}

In this study, we have investigated the spectroscopic parameters of the exotic vector $X(4630)$ by means of the thermal QCD sum rule at non-zero temperature. We calculated the mass and decay constant of $X(4630)$ at $T\neq0$ using the two-point QCD sum rule approach up to the sixth order of the operator dimension by including non-perturbative contributions. We determined the working regions of the auxiliary parameters entering the thermal QCD sum rules. Our numerical results showed that the mass and decay constant of $X(4630)$ resonance stayed approximately non-changed up to the $T/T_{c}\simeq0.5$ and $T/T_{c}\simeq0.3$, respectively. After these points, they decreased with the increase in temperature. Near the deconfinement temperature, the mass and decay constant decrease by $9.8\%$ and $60\%$ compared to their vacuum values, respectively. The decreasing of physical quantities near the deconfinement temperature can be considered as a sign of transition to QGP and these remarkable decreases in the values of the mass and the decay constant with temperature are in good agreement as behaviorally with other studies \cite{Veliev, Sungu1,  Sungu2, Sungu2, Sungu3, Montana, Montesinos} in the literature. 

Also, we observed that our results on the mass and decay constant at zero temperature are consistent with the results of LHCb experiments and other studies. We hope that our results in this work will help to the better interpretation of future heavy ion collision experiments data.

\appendix

\section{The temperature-dependent spectral density for $X(4630)$ resonance}\label{sec:App}

This appendix presents the explicit forms of different dimensional components of the temperature-dependent spectral density for $X(4630)$ resonance. The temperature-dependent spectral density $\rho^{\mathrm{OPE}}(s,T)$ is composed of these components as 
\begin{eqnarray}
\rho^{\mathrm{OPE}}(s,T)&=&\rho_{pert}^{\mathrm{OPE}}(s,T)+\rho_{dim3}^{\mathrm{OPE}}(s,T)+\rho_{dim4}^{\mathrm{OPE}}(s,T)\nonumber\\
&+&\rho_{dim5}^{\mathrm{OPE}}(s,T)+\rho_{dim6}^{\mathrm{OPE}}(s,T).
\end{eqnarray}

\begin{widetext}

The explicit forms of above components of $\rho^{\mathrm{OPE}}(s,T)$ are given as follows:

\begin{eqnarray}
\nonumber \rho_{pert}^{\mathrm{OPE}}(s,T)&=&\int_{0}^{1}dx_{1}\int_{0}^{1-x_{1}}~dx_{2}\frac{\Big(\kappa s x_{1}x_{2}-\alpha m_{c}^2\left(x_{1}+x_{2}\right)\Big)^2}{1536 \kappa^2 \alpha^8 \pi ^6 } \lbrace \alpha^2 m_{c}^4 x_{1} x_{2} \left(x_{1}+x_{2}\right) \Big(5x_{1}^2+5x_{2}\xi +x_{1}(8x_{2}-5)\Big) \nonumber\\
&-&\kappa^3 s x_{1} x_{2}\Big(12m_{s}^2\beta^2 -35s x_{1}^2 x_{2}^2 \Big)
-2\kappa \alpha m_{c}^2  \Big(6m_{s}^2 \beta^2(5x_{1}\chi+x_{2}(4x_{1}-5)+5x_{2})\nonumber\\
&+&sx_{1}^2 x_{2}^2 \left(14x_{1}\chi+x_{2}(27x_{1}-14)+14x_{2}^2\right)\Big)\rbrace
\end{eqnarray}

\begin{eqnarray}
\nonumber \rho_{dim3}^{\mathrm{OPE}}(s,T)&=&\int_{0}^{1}dx_{1}\int_{0}^{1-x_{1}}~dx_{2}\frac{m_{s}\langle\bar{s}s\rangle}{16 \alpha^6 \pi ^4 } \lbrace -\kappa^{3} s^2 x_{1}^2 x_{2}^2 \left(x_{1}^2+ x_{2}\xi-x_{1}\left(1+14x_{2}\right)\right)\nonumber\\
&+&3\alpha^2 m_{c}^4\left(x_{1}+x_{2}\right) \Big(x_{1}^4+x_{2}^2\xi^{2}+x_{1}^3 (3x_{2}-2)+ x_{1} x_{2}\xi(3x_{2}-2)+x_{1}^2(1+5 x_{2}\xi)\Big)\nonumber\\
&-& \kappa \alpha s m_{c}^2 x_{1} x_{2} \Big(2x_{1}^4+2\xi^{2} x_{2}^2+x_{1}^3 (19x_{2}-4)+ \xi x_{1} x_{2}(19x_{2}-4)+x_{1}^2(2+x_{2}(37x_{2}-23)\Big)\rbrace
\end{eqnarray}
\begin{eqnarray}
\nonumber \rho_{dim4}^{\mathrm{OPE}}(s,T)&=&\int_{0}^{1}dx_{1}\int_{0}^{1-x_{1}}~dx_{2}\frac{1}{9216 \alpha^6 \kappa^{2} \pi ^6 } \lbrace 768 \kappa^2 \pi^2 \langle u\Theta^{f}u\rangle x_{1} x_{2} \Big(30 \kappa^3 s^2 x_{1}^2 x_{2}^2+\alpha^2 m_{c}^4 (x_{1}+x_{2}) (x_{1}^2+\chi x_{2}+x_{1}(3x_{2}-1))\nonumber\\
&-& \alpha \kappa m_{c}^2 s x_{1} x_{2}\left(25 x_{1}^2+25\chi x_{2}
+x_{1}(51x_{2}-25)\right)\Big)+ g_{s}^2 \langle u\Theta^{g}u\rangle \Big(-120\kappa^4 s^2 x_{1}^3 x_{2}^3 (x_{1}+x_{2})+48\alpha^3 \kappa^{2}m_{c}^3 m_{s}(x_{1}+x_{2})^3\nonumber\\
&-&48\alpha^2 \kappa^{3}m_{c} m_{s} s x_{1} x_{2}\left(x_{1}^2+10 x_{1} x_{2}+x_{2}^2\right)+3\alpha^2 m_{c}^4 x_{1} x_{2}(x_{1}+x_{2}) (2x_{1}^2+2\chi x_{2}+x_{1}(3x_{2}-2)) \nonumber\\
&\times &\left(9x_{1}^2+9\chi x_{2}+x_{1}(17x_{2}-9)\right)+\alpha \kappa m_{c}^2 s x_{1}^2 x_{2}^2 (18x_{1}^4+18 \chi^2 x_{2}^2+x_{1}^3 (101x_{2}-36)+\chi x_{1} x_{2}(101x_{2}-36)\nonumber\\
&+&x_{1}^2(x_{2}(167x_{2}-137)+18))\Big)-\langle \frac{\alpha_s G^2}{\pi}\rangle \Big(240\kappa^4 s^2 x_{1}^3 x_{2}^3 (x_{1}+x_{2})+144\alpha^3 \kappa^{2}m_{c}^3 m_{s}(x_{1}+x_{2})(x_{1}^2-6x_{1}x_{2}+x_{2}^2)\nonumber\\
&-&48\alpha^2 \kappa^{3}m_{c} m_{s} s x_{1} x_{2}\left(3x_{1}^2-26 x_{1} x_{2}+3x_{2}^2\right)+\alpha^2 m_{c}^4 x_{1} x_{2}(x_{1}+x_{2})(34x_{1}^4+2\chi x_{2}^2(17x_{2}-33)+ x_{1}^3 (183x_{2}-100)\nonumber\\
&+&3x_{1}x_{2}(x_{2}(61x_{2}-113)+44)+x_{1}^2(x_{2}(277x_{2}-339)+66)))+\alpha \kappa m_{c}^2(24\alpha^2 m_{s}^2(3x_{1}^4+3 \chi x_{2}^3-x_{1}x_{2}^3\nonumber\\
&-&x_{1}^3 (x_{2}+3))-sx_{1}^2x_{2}^2(186x_{1}^4+6 \chi x_{2}^2(31x_{2}-43)+x_{1}^3 (915x_{2}-444)+x_{1}x_{2}(5x_{2}(183x_{2}-299)+516)\nonumber\\
&+&x_{1}^2(1437x_{2}-1495)+258)))\Big)\rbrace
\end{eqnarray}
\begin{eqnarray}
\nonumber \rho_{dim5}^{\mathrm{OPE}}(s,T)&=&\frac{m_0^2 m_{c} m_{s}\langle\overline{s}s\rangle}{384\pi^4}\int_{0}^{1}dx_{1}\Big(12m_{c}+m_{s}(6x_{1}-2)\Big)-\int_{0}^{1}dx_{1}\int_{0}^{1-x_{1}}~dx_{2}\frac{\kappa m_0^2\langle\overline{s}s\rangle}{192 \alpha^5 \pi ^4 } \nonumber\\
&\times &\lbrace -3\alpha \kappa m_{c} s x_{1}\left(2x_{1}-x_{2}\right)x_{2} -16\kappa^2 m_{s}s x_{1}^2 x_{2}^2+3\alpha^2 m_{c}^3\left(2x_{1}-x_{2}\right)(x_{1}+x_{2})\nonumber\\
&+&6\alpha \kappa  m_{c}^2 m_{s} x_{1} x_{2}\left(x_{1}+x_{2}\right)\rbrace
\end{eqnarray}
\begin{eqnarray}
\nonumber \rho_{dim6}^{\mathrm{OPE}}(s,T)&=&\frac{\langle\overline{s}s\rangle^2}{2592\pi^4}\int_{0}^{1}dx_{1}\Big(2g_{s}^2m_{c}m_{s}(3x_{1}-1)+27\pi^2(8m_{c}^2 +4m_{c} m_{s}(1-3x_{1})+5m_{s}^2(x_{1}-1)x_{1}\Big)\nonumber\\
&+&\int_{0}^{1}dx_{1}\int_{0}^{1-x_{1}}~dx_{2}\frac{g_{s}^2\kappa^2 \langle\overline{s}s\rangle^2 x_{1}x_{2}}{648 \alpha^5 \pi ^4 } \Big(-8\kappa s x_{1}x_{2}+3\alpha m_{c}^2 \left(x_{1}+x_{2}\right)\Big)
\end{eqnarray}

In these expressions, we used the below-defined notations: 

\begin{eqnarray}
\alpha&=&\Big(x_1^2+x_1(x_2-1)+x_2(x_2-1)\Big),  \nonumber \\
\beta&=&\Big(x_2^2+x_1(x_1-1)+x_2(x_1-1)\Big),  \nonumber \\
\kappa&=&(x_1+x_2-1),  \nonumber \\
\chi&=& (x_1-1),\nonumber \\
\xi&=&(x_2-1)
\end{eqnarray}

Also, $x_{1}$, $x_{2}$ and $x_{3}$ are Feynman integral parameters. 
\end{widetext}
%

%
%
\end{document}